\documentclass{iaus}

\usepackage{psfig,epsfig,graphicx}
\usepackage{amsmath}
\usepackage{natbib}

\title[Contribution of microlensing to quasar variability]{Contribution of microlensing to X-ray variability of
distant QSOs}

\author[Zakharov, Popovi\'c \& Jovanovi\'c]%
{Alexander F. Zakharov$^{1,2}$%
  \thanks{Present address: Institute of Theoretical and
Experimental Physics,
           25, B.Cheremushkinskaya st., Moscow, 117259, Russia.},
Luka \v C. Popovi\'c$^{3,4}$\break \and Predrag  Jovanovi\'c$^3$}

\affiliation{$^1$ Institute of Theoretical and
Experimental Physics,
           25, B.Cheremushkinskaya st., Moscow, 117259, Russia email:zakharov@itep.ru\\[\affilskip]
$^2$Astro Space
Centre of Lebedev Physics Institute, Moscow, Russia \\[\affilskip]
$^3$Astronomical Observatory, Volgina 7, 11160 Beograd, Serbia email:lpopovic@aob.bg.ac.yu\\[\affilskip]
$^4$ Astrophysikalisches Institut Potsdam, An der Sternwarte 16,
14482 Potsdam, Germany, email:lpopovic@aip.de }

\pubyear{2004}
\volume{225}
\pagerange{1--8}
\date{?? and in revised form ??}
\jname{Impact of Gravitational Lensing on Cosmology}
\editors{Mellier, Y. \& Meylan,G. eds.}
\begin{document}

\maketitle

\begin{abstract}
We consider a contribution of microlensing to the X-ray
variability of high-redshifted QSOs.
Cosmologically distributed gravitational microlenses could be
localized in galaxies (or even in bulge or halo of  gravitational
macrolenses) or could be distributed in a uniform way. We have
analyzed both cases of such distributions. We found that the
optical depth for gravitational microlensing caused by
cosmologically distributed deflectors could be significant and
could reach $10^{-2} - 0.1$ at $z\sim 2$.  This means that
cosmologically distributed deflectors may contribute
significantlly  to the X-ray variability of high-redshifted QSOs
($z>2$). Considering that the upper limit of the optical depth
($\tau\sim 0.1$) corresponds to the case where dark matter forms
cosmologically distributed deflectors,   observations of the X-ray
variations of unlensed QSOs can be used for the estimation of the
dark matter fraction of microlenses.

\end{abstract}

\firstsection 

\section{Introduction}

 X-ray flux variability has long been known to be a common property of
active galactic nuclei (AGNs), e.g. Ariel 5 and HEAO 1 first
revealed long-term (days to years) variability in AGNs  and by
uninterrupted observations of EXOSAT  rapid (thousands of seconds)
variability was also established as common in these sources (see,
for example reviews by \cite{Mush93,Ulrich97} and references
therein). X-ray flux variations are observed on timescales from
$\sim$1000 s to years, and amplitude variations of up to an order
of magnitude are in the $\sim$ 0.1 - 10 keV spectral band.
Recently, \cite{Man02}
 analyzed the variability of a sample of 156 radio-quiet quasars
taken from the ROSAT archive, considering the trends in variability of the
amplitude with luminosity and with redshift. They found that there
was evidences for a growth in AGN X-ray variability amplitude
towards high redshift ($z$) in the sense that AGNs of the same
X-ray luminosity were more variable at $z>2$. They explained the
$\sigma$ {\it vs.} $z$ trend assuming that the high-redshifted
AGNs accreted at a larger fraction of the Eddington limit than
the low-redshifted ones.

On the other hand, the contribution of microlensing to AGN
variability was considered in many papers (see e.g.
\cite{Hawk93,Hawk02, Wamb01,Wamb01b,Zakh97}, and references
therein). Moreover, recently X-ray microlensing of AGN has been
considered
\citep{Popov01a,Tak01,Chart02a,Popovic03,Popovic03b,Dai03}. Taking
into account that the X-rays of AGNs are generated in the
innermost and very compact region of an accretion disc, the X-ray
radiation in the continuum as well as in a line can be strongly
affected by microlensing \citep{Popovic03}.\footnote{Simulations
of X-ray line profiles are presented in a number of papers, see,
for example,
\cite{Zak_rep02,Zak_rep02a,Zak_rep02_xeus,Zak_rep02_Gamma,Zak_rep03_Lom,Zak_rep03_ASR,Zak_rep03_Su,Zak_rep03_Sakh}
and references therein, in particular \cite{ZKLR02} showed that an
information about magnetic filed may be extracted from X-ray line
shape analysis; \cite{Zak_rep03_AA} discussed signatures of X-ray
line shapes for highly inclined accretion disks, \cite{ZMB03}
calculated shapes of spectral lines for non-flat accretion flows.}
 Recent
observations of three lens systems seem to support this idea
\citep{Osh01,Chart02a,Cha04,Dai03}.  \cite{Popovic03,Popovic03b}  showed that
objects in a foreground galaxy with very small masses can cause
strong changes in the X-ray line profile.  This  fact may
indicate  that the observational probability of X-ray variation
due to microlensing events is higher  than in
the UV and optical radiation of AGNs. It is connected with
the fact that typical sizes of X-ray emission regions are much
smaller than typical sizes of those producing optical and UV bands.
Typical optical and UV emission region sizes could be
comparable or even larger than Einstein radii of microlenses and
therefore microlenses magnify  only a small part of the region emitting in
the optical or UV band (see e.g. \cite{Pop01b,Aba02}, for UV and
optical spectral
line
region). This is reason that it could be a
very tiny effect from an observer point of view, in spite of  this fact recently
\cite{Richards04}
observed microlensing of the C IV line in SDSS J1004+4112.

Microlenses in quasar bulge/halo give a small contribution into
optical depth, therefore it would be reasonable to evaluate a
contribution  from cosmological distribution of microlenses
\citep{ZPJ03,ZPJ04}.

\section{Cosmological distribution of microlenses}

To estimate the optical depth we will use the point size source
approximation for an emitting region of X-ray radiation. It means
that the size of emitting region  is smaller than this
Einstein --
Chwolson radius. This approximation is used commonly to
investigate microlensing in optical and UV bands. The typical
 Einstein -- Chwolson radius of a lens can be expressed in
the following way \citep{Wamb01}
\begin {eqnarray}
r_{\rm EC}=\sqrt{\frac{4GM}{c^2}\frac{D_s D_{ls}}{D_l}} \sim 4
\times 10^{16} \sqrt{M/M_\odot}~{\rm cm},
 \label{eq_cosmol_l1}
\end {eqnarray}
where "typical" lens and source redshift of $z \sim 0.5$ and $z
\sim 2$ were chosen, $M$ is the lens mass, $D_l$, $D_s$ and
$D_{ls}$ are angular diameter distances between an observer and a
lens, observer and source, lens and source respectively. A typical
quasar size is parameterized in units of $10^{15}$ cm
\citep{Wamb01}. Since the point size source approximation for an
emitting region is reasonable for optical and for UV bands, and as
it is generally adopted that X-ray radiation is formed in the
inner parts of accretion disks we can use this approximation for
X-ray sources. However, let us present simple estimates. The
relevant length scale for microlensing in the source plane for
this sample
\begin {eqnarray}
R_{\rm EC}=r_{\rm EC}\frac{D_s}{D_l} \sim 1 \times 10^{17}~{\rm
cm}.
 \label{eq_cosmol_l2}
\end {eqnarray}

Even if we consider a supermassive black hole in the center of the
quasar $M_{\rm SMBH}= 10^{9} M_\odot$, then its Schwarzschild
radius is $r_g=3\times 10^{14}$ cm and assuming that the emission
region for the X-ray radiation is located near the black hole
$r_{\rm emission} < 100\,r_g =3\times 10^{16}$~cm, we obtain that
$r_{\rm emission} < R_{\rm EC}$, therefore the point size source
approximation can be adopted for the X-ray emitting
region.\footnote{For example, \cite{Chart02a} found evidence for
X-ray microlensing in the gravitationally lensed quasar MG
J0414+0534 ($z=2.639$), where according to their estimates $M_{\rm
SMBH}$ is in the range $3.6 \times 10^{6}(\beta/0.2)^2$ and $1.1
\times 10^{7}(\beta/0.2)^2 M_\odot$ ($\beta \sim 1$).
Therefore a typical emission region is much smaller than the
Einstein -- Chwolson radius $R_{\rm EC}$, since following
\cite{Chart02a} one could assume that the emitting region corresponds
to $(10-1000)\,r_g$ or $\sim 1.5 \times 10^{14} - 1.5 \times
10^{16}$~cm for a $10^{8}M_\odot$ black hole.}

To  evaluate the optical depth, we assume  a source located at
redshift $z$.
The expression for optical depth  has been taken  from
\cite{TOG84,Fuk91}
\begin {eqnarray}
\tau^p_L= \frac{3}{2}\frac{\Omega_L}{\lambda(z)}
\int_0^z dw
\frac{(1+w)^3[\lambda(z)-\lambda(w)]\lambda(w)}
       {\sqrt{\Omega_0(1+w)^3+\Omega_\Lambda}},
       \label{eq_cosmol2}
\end {eqnarray}
where $\Omega_L$ is the matter fraction in compact lenses,
\begin {eqnarray}
\lambda(z)= \int_0^z
\frac{dw}{(1+w)^2\sqrt{\Omega_0(1+w)^3+\Omega_\Lambda}},
\label{eq_cosmol3}
\end {eqnarray}
is the affine distance (in units of $cH^{-1}_0)$.

\begin{figure}[t!]
\begin{center}
\includegraphics[width=7.5cm]{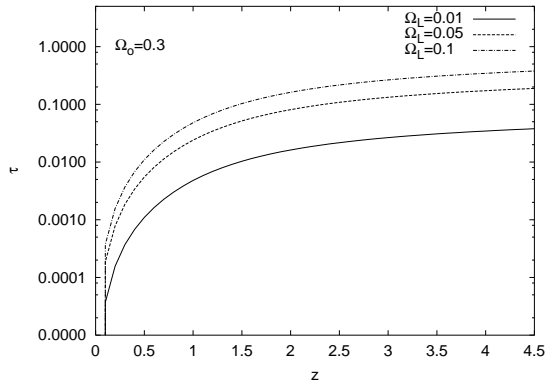}

\end{center} \caption{The calculated optical depth as a function of
redshift for different values of $\Omega_L$ and $\Omega_0$.}
 \label{fig1}
\end{figure}

We will use some realistic cosmological parameters to evaluate the
integral (\ref{eq_cosmol2}). According to the cosmological SN
(Supernova) Ia data
 and  cosmic microwave background (CMB) anisotropy one can take
$\Omega_\Lambda \approx 0.7,
\Omega_0  \approx 0.3$
\citep{Per99}. Recent CMB anisotropy
observations by the WMAP satellite team have confirmed important
aspects of the current standard cosmological model,  the
WMAP team determined  $\Omega_\Lambda \approx 0.73, \Omega_0
\approx 0.27$ \citep{Ben03,Sper03} for the "best" fit of
cosmological parameters.  If we assume that
microlensing is caused by stars we have to  take into account
cosmological constraints  on baryon density.  Big Bang
Nucleosynthesis (BBN) calculations together with observational
data about the abundance of $^2$D give the following constraints on
the cosmic baryon density \citep{
Turner02}
\begin {eqnarray}
\Omega_b h^2 = 0.02 \pm 0.002,
\label{eq_cosmol4}
\end {eqnarray}
 taking into account the Hubble constant estimation $h=0.72\pm
0.08$ \citep{Freedman01}. An analysis of recent WMAP data on CMB
anisotropy gives as the best fit \citep{Sper03}
\begin {eqnarray}
\Omega_b h^2 = 0.0224 \pm 0.0009, \label{eq_cosmol7}
\end {eqnarray}
which is very close  to the BBN constraints, but with much
smaller error bars.

 Therefore, the cases with $\Omega_0=0.3$ and $\Omega_L =
0.05$ ($\Omega_L = 0.01$) can be adopted as realistic.
Here we assume that almost all baryon
matter
can form microlenses ($\Omega_L = 0.05$), or, alternatively, that about
25\% of
baryon matter forms such microlenses ($\Omega_L = 0.01$)).

\section{Discussion and results}

In Fig.  we show the optical depth of cosmologically distributed
microlenses assuming three cases of cosmoligically distributed
microlenses: i) small fraction of baryonic matter (25 \%) forms
microlenses ($\Omega_L = 0.01$); ii) almost all baryonic matter
forms microlenses ($\Omega_L = 0.05$); iii) about 30\% of
non-baryonic (dark) matter forms microlenses ($\Omega_L = 0.1$).

As  was mentioned earlier by \cite{Popovic03,Popovic03b} the
probability of microlensing by stars or other compact objects in
halos and bulges of quasars is very low (about $10^{-4} -
10^{-3}$). However, as one can see from Fig.~1, for cosmologically
distributed microlenses it could reach $10^{-2} - 0.1$ at $z\sim
2$. The upper limit $\tau \sim 0.1$ corresponds to the case where
compact dark matter forms cosmologically distributed microlenses.
As one can see from Fig. 1, in this case the optical depth for the
considered value of $\Omega_0$  is around 0.1  for
 $z > 2$. This indicates that   such a phenomenon could be  observed
frequently, but only for
distant sources ($z \sim 2$).

To investigate  distortions of spectral line shapes due to
microlensing \citep{Popovic03,Popovic03b} the most real candidates
are multiply imaged quasars. However, these cases the simple
point-like microlens model may  not be very good approximation
\citep{Wamb01,Wamb01b} and one should use a numerical approach,
such as the MICROLENS ray tracing program, developed by J.
Wambsganss 
or  some analytical approach
for magnification near caustic curves like folds
\citep{Schneider92a,Fluke99} or near singular caustic points like
cusps \citep{Schneider92,Mao92,Zakharov95,Zakharov97,Zakharov99}
as was realized by \cite{Yonehara01}.

If  we believe in the  observational arguments of \cite{Hawk02} that the
variability of a significant fraction of distant quasars is caused by
microlensing,  the analysis of the properties of X-ray line shapes
due to microlensing \citep{Popovic03} is a powerful tool to
confirm or rule out Hawkins' (2002) conclusions.

As it was mentioned, the probability that the shape of the  Fe $K\alpha$
line is distorted (or amplified) is highest in gravitationally lensed
systems.
 Actually, this phenomena was
discovered by \cite{Osh01,Dai03,Chart02a,Cha02b,Cha04} who found
evidencees for such an effect for   QSO H1413+117 (the Cloverleaf,
$z=2.56$), QSO~2237+0305 (the Einstein Cross, $z=1.695$), MG
J0414+0534 ($z=2.64$) and possibly for BAL QSO~08279+5255 ($
z=3.91$).
 One could say that it is
natural that the discovery of X-ray microlensing was made for this
quasar, since the Einstein Cross QSO~2237+0305 is the
most "popular" object to search for microlensing, because the
first cosmological microlensing  phenomenon was found by
\cite{Irwin89} in this object and  several groups have been
monitoring the quasar QSO~2237+0305 to find evidence for
microlensing.  Microlensing has been suggested for the quasar
MG J0414+0534 \citep{Angonin99} and for the quasar QSO H1413+117
\citep{Remy96}. Therefore,
in future a chance may be to find X-ray microlensing for other
gravitationally lensed systems that have  signatures of
microlensing in the optical and radio bands. Moreover, considering the
sizes of the sources of
X-ray radiation, the variability in the X-ray range during microlensing
event
should be more prominent than in the optical and UV.
{ Consequently, gravitational microlensing in the X-ray band
 is a powerful tool for
 dark matter investigations, as the upper limit of optical
depth ($\tau\sim 0.1$) corresponds to the case where dark matter forms
cosmologically distributed deflectors.

\section*{Conclusions}

From our calculations we can conclude \citep{ZPJ03,ZPJ04}:

i) The optical depth for  cosmologically distributed deflectors could
 be $\sim 10^{-2}-0.1$ at $z\sim 2$ and might contribute  significantly
 to the
X-ray variability of high-redshift QSOs. The value $\tau\sim 0.1$
corresponds to the
case where compact dark matter forms cosmologically distributed
microlenses.

ii) The optical depth for  cosmologically distributed deflectors
($\tau_L^p$)  is higher for $z>2$ and increases slowly beyond $z=2$.
 This indicates that the contribution of microlensing on the
X-ray variability of QSOs with redshift $z>2$ may be significant,
and also  that this contribution could  be nearly constant for
high-redshift QSOs. This is in good agreement with the fact that
AGNs of the same X-ray luminosity are more variable at $z>2$
\citep{Man02}.

 iii) Observations of X-ray variations of unlensed QSOs can be used for estimations of matter fraction of microlenses.
 The rate of
 microlensing can be used for estimates of the cosmological density of
microlenses, and consequently  the fraction of dark
matter microlenses, but the durations of microlensing events could be used
for gravitational microlens mass estimations.

\begin{acknowledgements}

This work was supported in part by  the Ministry of Science,
Technologies and Development of Serbia through the project
"Astrophysical Spectroscopy of Extragalactic Objects" (L\v CP \&
PJ) and the Alexander von Humboldt Foundation through the program
for foreign scholars (L\v CP).

Authors are grateful to Y. Mellier \& G. Meylan for their
attention to this contribution.

\end{acknowledgements}

\end{document}